\documentclass{article}

\usepackage[english]{babel}

\usepackage[letterpaper,top=2cm,bottom=2cm,left=3cm,right=3cm,marginparwidth=1.75cm]{geometry}

\usepackage{amsmath}
\usepackage{graphicx}
\usepackage{authblk}
\usepackage[colorlinks=true, allcolors=blue]{hyperref}
\usepackage[version=3]{mhchem} 
\usepackage[capitalise]{cleveref}
\usepackage{siunitx}

\title{Electrochemical removal of amphoteric ions}
\author[a\footnote{Equally contributed}]{Amit N. Shocron}
\author[a$^*$]{Eric N. Guyes} 
\author[b]{P. Maarten Biesheuvel}
\author[c]{Huub H.M. Rijnaarts}
\author[a,d,e\footnote{To whom correspondence should be addressed. E-mail: mesuss@me.technion.ac.il, jouke.dykstra@wur.nl}]{Matthew E. Suss}
\author[c$^\dagger$]{Jouke E. Dykstra}

\affil[a]{Faculty of Mechanical Engineering, Technion - Israel Institute of Technology, Haifa, Israel}
\affil[b]{Wetsus, European Centre of Excellence for Sustainable Water Technology, Oostergoweg 9, 8911 MA Leeuwarden, The Netherlands}
\affil[c]{Environmental Technology, Wageningen University, Bornse Weilanden 9, 6708 WG Wageningen, The Netherlands}
\affil[d]{Grand Technion Energy Program, Technion - Israel Institute of Technology, Haifa, Israel}
\affil[e]{Wolfson Department of Chemical Engineering, Technion - Israel Institute of Technology, Haifa, Israel}

\begin{document}
\maketitle

\begin{abstract}
Several harmful or valuable ionic species present in sea, brackish and wastewaters are amphoteric, and thus their properties depend on the local water pH. Effective removal of these species can be challenging by conventional membrane technologies, necessitating chemical dosing of the feedwater to adjust its pH. Capacitive deionization (CDI) is an emerging membraneless technique for water treatment and desalination, based on electrosorption of salt ions into charging microporous electrodes. CDI cells show strong internally-generated pH variations during operation, and thus CDI can potentially remove amphoteric species without chemical dosing. However, development of this technique is inhibited by the complexities inherent to coupling of pH dynamics and amphoteric ion properties in a charging CDI cell. Here, we present a novel theoretical framework predicting the electrosorption of amphoteric species in flow-through electrode CDI cells. We demonstrate that such a model enables deep insight into factors affecting amphoteric species electrosorption, and conclude that important design rules for such systems are highly counter-intuitive. For example, we show both theoretically and experimentally that for boron removal the anode should be placed upstream, which runs counter to accepted wisdom in the CDI field. Overall, we show that to achieve target separations relying on coupled, complex phenomena, such as in the removal of amphoteric species, a theoretical CDI model is essential. 
\end{abstract}

\section{Introduction}

Global fresh water scarcity is increasing due to population growth, increased water consumption per capita, and shrinking freshwater bodies due to climate change and over-extraction.\cite{Montgomery2007,Richter2013,Damania2017} About 4 billion people are subject to severe water scarcity for at least one month per year,\cite{Mekonnen2016} and increasing parts of the global population are predicted to face chronic water scarcity.\cite{Schewe2014} Consequently, there has been increasing demand for efficient water treatment and desalination technologies over the past few decades.\cite{Elimelech2011,Jones2019} Water treatment also provides an opportunity for recovery of valuable elements from feedwater.\cite{Cordell2009,Nir2018}

Commonly used technologies for water treatment and desalination are pressure-driven membrane-based separation technologies, such as reverse-osmosis (RO),\cite{Fritzmann2007,Wenten2016,Jones2019} nanofiltration and ultrafiltration. \cite{Rautenbach1997,Zeman2017} During operation, mechanical energy is invested in pumping and pressurizing feedwater through the membrane, while ions and other compounds can be retained by the membrane. Two important mechanisms for the rejection of ions and molecules by membranes are size-expulsion,\cite{Wang1995,Garba1999}, and charge repulsion,\cite{Wang1995,Garba1999,Childress2000}. An ion's volume includes its hydration shell, which often plays an important role in rejection.

The charge of several common pollutants and valuable solutes depends on the solution pH, and such species are characterized as amphoteric. Arsenic acid, As(V), and arsenous acid, As(III), are toxic amphoteric ions, and therefore should be removed during drinking water treatment.\cite{Zaldivar1974,Tseng1977,Farias2003} Boron is an amphoteric ion which is considered toxic when at high concentrations in water,\cite{Lee1978,Gupta1985} and which can be detrimental to plant growth.\cite{USEPA1975,Moss2003,Bick2005} Phosphate and ammonia should be removed from wastewater, as these amphoteric ions can negatively affect the surface water quality when present at high concentrations.\cite{Montgomery2007,Penate2012,Jin2014,Mayer2016}. They are also desirable to recover during water treatment, \cite{Cordell2009,Montoya2015,Mayer2016,Nir2018,Wang2019} as they are important nutrients.\cite{Gruber2008} Acetate is a small organic acid and amphoteric, and is often removed and recovered during sugar production.\cite{Wooley1998,Ahsan2014} As the hydration shell of amphoteric ions varies with their charge, the rejection of amphoteric species by membranes is generally pH-dependent, and may be highly challenging under certain conditions.\cite{Tu2010}

\begin{figure}[!hbtp]
\centering
	\includegraphics[width=0.9\linewidth]{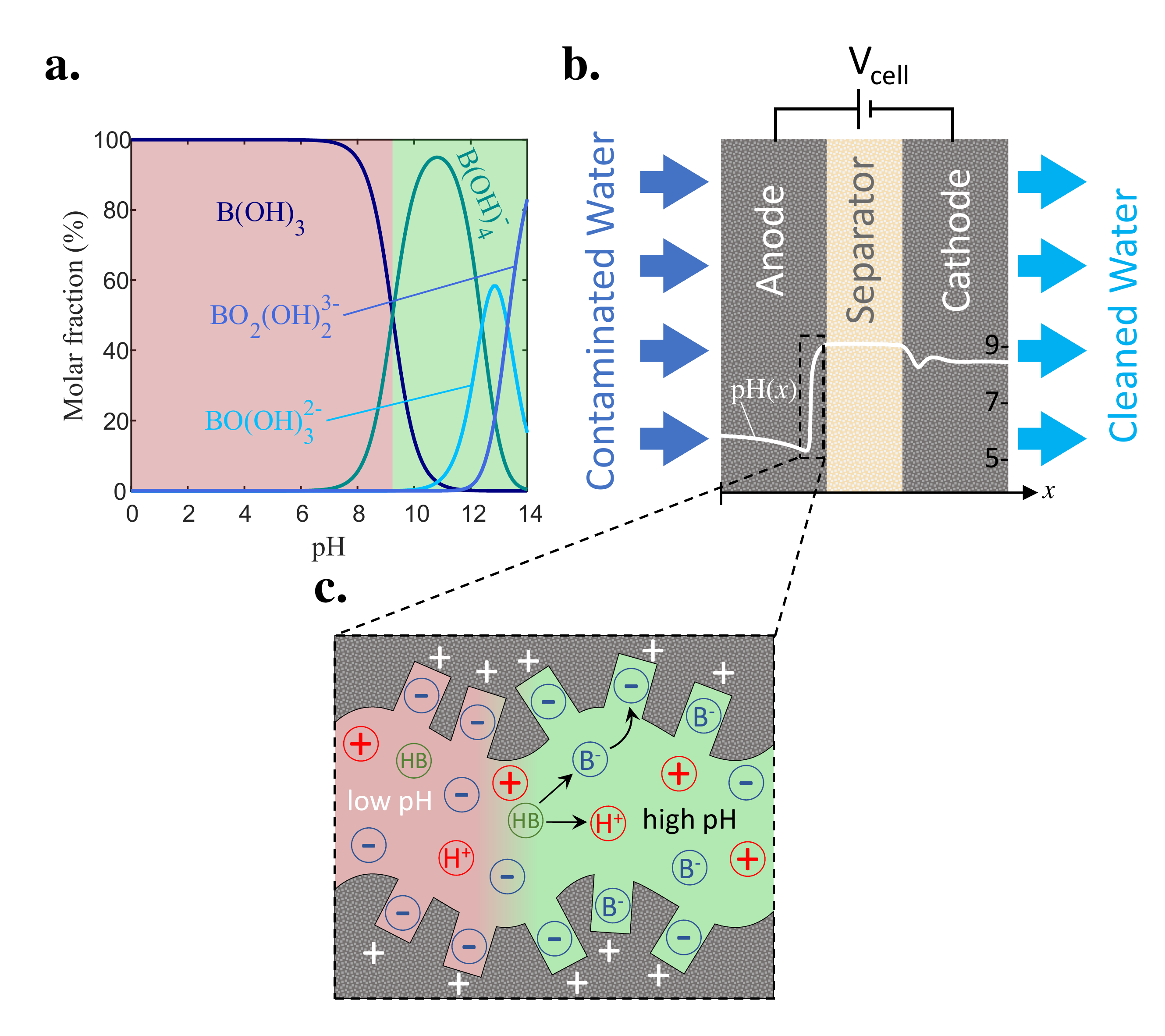}
	\caption{Schematic demonstrating boron removal by a CDI cell. a. Boron is amphoteric in water, and plotted is the molar fraction of the various boron-containing species as a function of water pH. b. A CDI cell with an anode placed upstream, and a snapshot of the developed pH profile within the electrodes. c. A snapshot of ion and charge distributions near the anode/separator interface, showing boric acid dissociation and electrosorption.}
	\label{fig:AmphSchematics}
\end{figure}

Boron is an example of an amphoteric ion that can be poorly rejected in membrane-based systems. Boron is present in protonated form, \ce{B(OH)3}, under standard pH conditions (pH of surface water is generally between 7 and 8), and can dissociate into \ce{B(OH)4-} and \ce{BO(OH)3^{2-}} at higher pH, see \cref{fig:AmphSchematics}a. Since \ce{B(OH)3} is not charged, the hydration of the molecule is weak, and \ce{B(OH)3} can pass through the pores of an RO membrane, resulting in a low removal rate. The removal is generally only around 50-60\% for feedwater with a pH below 8, but values ranging from 10-30\% have been reported as well.\cite{Magara1996,Prats2000,Redondo2003,Xu2010,Tu2011} To increase boron removal with RO, common practice is to pass desalinated water multiple times through an RO membrane.\cite{Glueckstern2003,Nadav2005,Greenlee2009} Another practice is to dose, after the first RO pass and before the second RO pass, a caustic solution to adjust the pH to higher values.\cite{Greenlee2009,Kabay2010,Rahmawati2012} This adjustment results in boron primarily appearing within \ce{B(OH)4-} and \ce{BO(OH)3^{2-}}, species which are blocked by the RO membrane due to their larger hydrated size.

Capacitive deionization (CDI) is an emerging membraneless electrochemical separation technology used for water treatment and desalination. Typical CDI cells employ a cyclic process of alternatingly charging and discharging a pair of nanoporous carbon electrodes. During the charging step, ions are removed from the feedwater by electrosorption into the electrodes, and during the discharge step the electrodes are regenerated and the ions are released into the brine. Several cell designs for CDI have been proposed, including flow-by electrode CDI, membrane CDI, and flow-through electrode (FTE) CDI,\cite{Suss2015} the latter schematically shown in \cref{fig:AmphSchematics}b. In FTE CDI, the solution flows through the electrodes,\cite{Suss2012,Suss2015} which allows the use of a thinner separator compared to other cell designs, enabling reduced cell resistance\cite{Dykstra2016} and faster desalination, but can lead to enhanced electrode degradation.\cite{Cohen2015,Zhang2019Comparison} Recent breakthroughs in CDI focus on its ability to remove ions selectively from polluted feedwaters,\cite{Seo2010,Zhao2012,Suss2017,Hawks2019Nitrate,Guyes2019,Mubita2019,Tsai2021,Guyes2021} and new electrode materials based on intercalation compounds or Faradaic reactions.\cite{Brousse2015,Lukatskaya2016,Srimuk2018,Bao2018,Shen2020}

During charging of a CDI cell with nanoporous carbon electrodes, large pH differences naturally develop in the cell, with pH values as high as $\sim$ 10 reported in the cathode and as low as $\sim$ 3 in the anode.\cite{Holubowitch2017,Landon2021} Several mechanisms have been proposed to explain these pH differences, including side-reactions such as oxygen reduction,\cite{He2016,Zhang2018a,Dykstra2017,Holubowitch2017,Landon2021} the difference in diffusion coefficients between ionic species present in solution, and electrosorption of \ce{H+} and \ce{OH-}.\cite{Holubowitch2017,Zhang2019Comparison,Dykstra2017} Thus far, one theoretical model was developed including pH effects in membrane CDI,\cite{Dykstra2017} with significant deviations between model and experiment, and no models have been developed for membraneless CDI, to our knowledge. Thus, pH effects and dynamics in CDI remain poorly understood, and the in-situ generated pH gradients in a CDI cell remain largely unexploited.

In the present work, we utilize these strong differences in local pH of charging CDI cells to ensure amphoteric species are charged and can be electrosorbed into the electrodes, see \cref{fig:AmphSchematics}b-c. Despite the presence of several experimental studies,\cite{Avraham2011,Kim2012,Fan2016,Fan2017,Huang2017,Wang2017,Fang2018,Sakar2019,Jiang2019,Fang2020} no theory has been developed to predict and guide the removal of amphoteric species by CDI. Given the complexities involving pH dynamics in CDI, a validated model is essential to unlock the enormous potential of CDI to remove amphoteric ions without chemical additives to adjust feed pH. We here provide such a theory, to our knowledge the first model coupling electrosorption into nanopores to local pH dynamics and acid-base equilibria, by a transient description of multi-component ion transport in porous electrodes, and amphoteric ions present in the feed. We show that our model predicts highly counter-intuitive design rules for removal of amphoteric species by CDI, which we validate experimentally using boron as a case study. In the future, the theoretical framework presented here can be adapted to any other amphoteric species, and can be used to improve the understanding of pH dynamics in electrochemical systems.

\section{Results and Discussion}
In this section, we employ the theoretical framework presented in the Materials and Methods section. This model captures local pH variations in the CDI cell, and the electrosorption of amphoteric species present in the feedwater. We consider an FTE CDI cell depicted in \cref{fig:TheorySchematics}a, with feedwater containing both a salt and an amphoteric species. To electrosorb the amphoteric species effectively, the species must be present in an ionic form while in the CDI cell. For example, ammonia is largely protonated and positively charged at neutral and low pH values ($\mathrm{p}K_\mathrm{a}$=9.25) , while boron is largely deprotonated and negatively charged at high pH conditions ($\mathrm{p}K_\mathrm{a}$=9.24), see \cref{fig:AmphSchematics}a. Thus, to remove boron, a high pH is required in the anode, and for ammonia removal, neutral or low pH is required in the cathode.

Here, we illustrate the theory by analyzing the electrosorption of boron as a case study, and compare the theoretical predictions to experimental data. As we do not expect the local pH in the cell to exceed the value of 10, we exclude the species \ce{BO(OH)_3^{2-}} and \ce{BO2(OH)2^{3-}} from our analysis (\cref{fig:AmphSchematics}a), and account only for \ce{B(OH)3}, which hence will be referred to as \ce{HB}, and \ce{B(OH)4-}, which will be denoted \ce{B-}. The feedwater we consider has a composition resembling the effluent from an RO system, so with relatively low salt concentration on the order of $\sim 1\ \rm{mM}$, a boron feed concentration of 0.37 mM, and a near-neutral feed pH $\left(6\le\rm{pH_F}\le 7\right)$.\cite{Hyung2006} 

\subsection{Electrode order affects local pH and boron removal}
The order of the electrodes in an FTE CDI cell may directly affect local pH values, and ultimately may be a central consideration for successful boron removal. Conventional wisdom is that during CDI cell charging, the solution in the cathode macropores becomes basic and in the anode acidic.\cite{He2016,Holubowitch2017} Naively, we may expect that at feed velocities characterized by Peclet number of order unity and above (${\rm Pe}\equiv v_{\mathrm{w}}l_{\mathrm{e}}/D_{\mathrm{s}}$, with $v_{\mathrm{w}}$ the superficial flow velocity, $l_{\mathrm{e}}$ the electrode thickness, and $D_{\mathrm{s}}$ the effective salt diffusion coefficient, see \cref{tab:general}), the cathode should be placed upstream so the high pH solution can be advected into the anode, allowing for boron electrosorption. As we will show, model and experimental results both prove that this expectation is not correct.

\begin{figure}[t]
	\centering
	\includegraphics[width=0.8\textwidth]{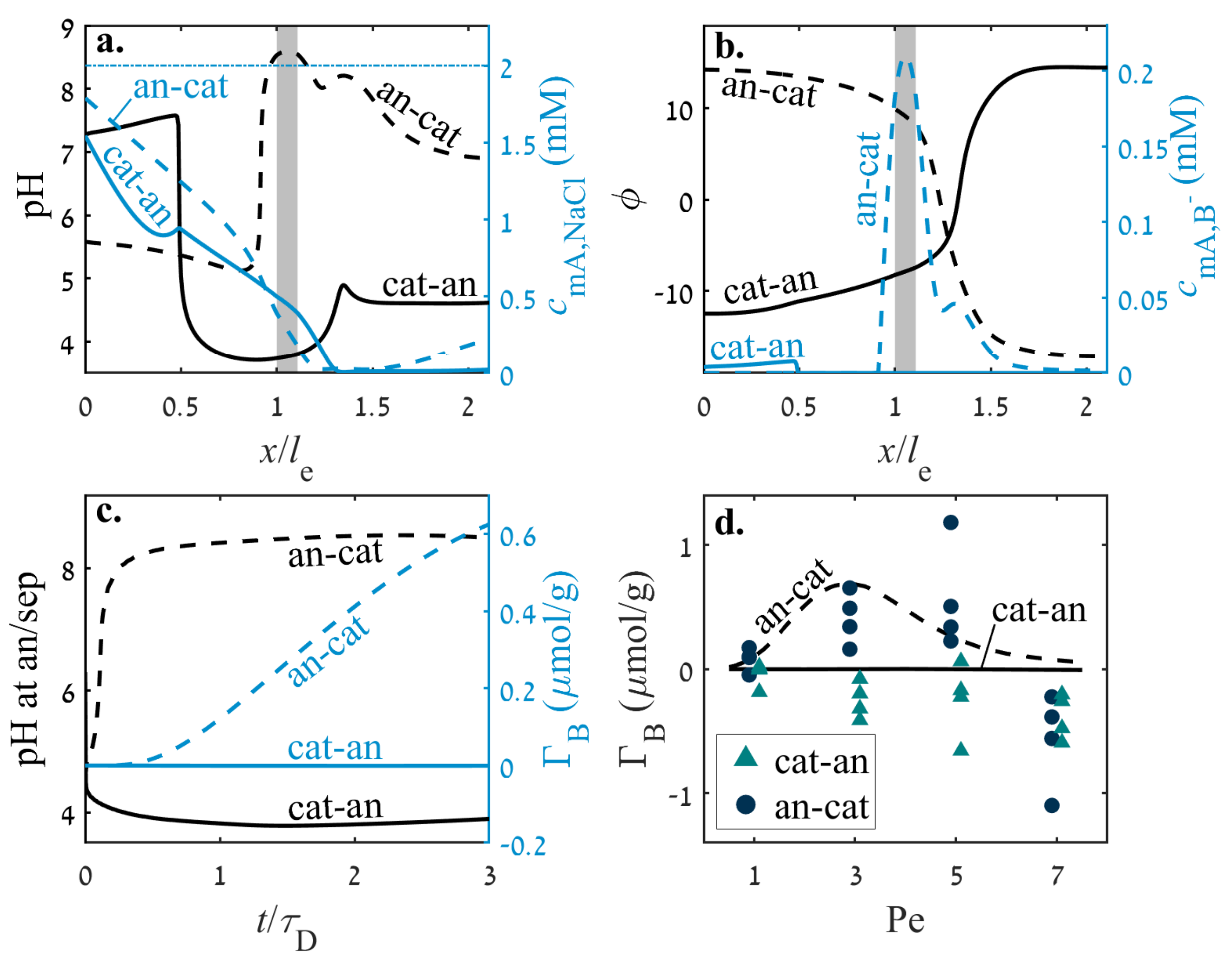}
	\caption{Predictions and experimental results for a FTE CDI cell with cathode placed upstream (cat-an, solid lines) or downstream (an-cat, dashed lines), for $c_{\mathrm{F}}$=2 mM, $c_{\mathrm{F,B}}$=0.37 mM and $V_{\mathrm{ch}}$=1.0 V, where in panels a-c Pe=3. a. Predicted pH (left y-axis, black lines) and macropore salt concentration (right y-axis, blue lines) profiles in the FTE CDI cell. b. Predicted scaled potential (left y-axis, black lines) and \ce{B-} macropore concentration (right y-axis, blue lines)  profiles. The profiles in panels a and b are snapshots at $t/\tau_D=2$, the gray-shaded rectangles represent the separator, and the horizontal dash-dotted blue line in panel a represents the feed salt concentration. c. Predicted pH at the anode/separator interface (left y-axis, black lines), and electrosorbed boron (right y-axis, blue lines), both as a function of dimensionless time. d. Predicted and measured boron boron electrosorption for an effluent collection time of 10 min as a function of Pe for cat-an (solid line, triangles) and an-cat (dashed line, circles) configurations.}
	\label{fig:ElectrodesOrder}
\end{figure}

In \cref{fig:ElectrodesOrder}a-c we compare the predicted salt, boron, potential, and pH profiles of two cell configurations operated with Pe=3 and a charging voltage of 1.0 V (\cref{tab:general,tab:electrodes}). In the first configuration, the cathode is placed upstream (cat-an, solid lines), and for the second the cathode is placed downstream (an-cat, dashed lines). \cref{fig:ElectrodesOrder}a shows predicted pH profiles (left y-axis, black lines) and macropore salt concentration (right y-axis, blue lines), defined by $c_{\mathrm{mA,NaCl}}\equiv \tfrac{1}{2}(c_{\mathrm{mA,Na^+}}+c_{\mathrm{mA,Cl^-}})$, both as a function of the location scaled by $l_{\mathrm{e}}$, and at $t/\tau_{\mathrm{D}}=2$, where $\tau_{\mathrm{D}}$ is the dimensionless time, defined as $l_{\mathrm{e}}^2/D_{\mathrm{s}}$. In \cref{fig:ElectrodesOrder}a we see that for both configurations the salt concentration decreases along the upstream electrode in flow direction, and is essentially completely depleted throughout most of the downstream electrode. For the cat-an configuration, the pH profile includes three distinct regions. First, in the upstream electrode (cathode) near the inlet, there is a relatively high pH of approximately 7.5. In the downstream half of the cathode, a sharp pH decrease to approximately 3.7 is observed. The latter feature is a novel observation to our knowledge, and runs contrary to conventional wisdom that the solution in the charging cathode is alkaline.\cite{Bouhadana2011,Holubowitch2017} Then, towards the downstream region of the anode pH rises again to values of approximately 4.6. For the an-cat configuration, the pH profile is changed significantly, with a pH of approximately 5.8 in the upstream part of the anode, followed by a sharp increase in pH close to the separator to about 8.6. The latter feature is again a novel observation and contrary to conventional wisdom that solution in the charging anode is acidic.\cite{Bouhadana2011,Holubowitch2017} Decreasing pH values are predicted along the rest of the cathode, reaching a value of about 6.9 at the effluent. 

Thus, results of \cref{fig:ElectrodesOrder}a indicate that the anode should be placed upstream for effective boron removal, counter to our naive expectation. To gain more insight into this counter-intuitive behavior, \cref{fig:ElectrodesOrder}b shows predicted profiles for potential, $\phi$, (left y-axis, black lines) and $c_{\mathrm{mA,\ce{B-}}}$ (right y-axis, blue lines) at $t/\tau_{\mathrm{D}}=2$.
The potential profiles in \cref{fig:ElectrodesOrder}b for both configurations are not symmetric about the cell midline, where in both cases $\phi$ changes sign in the downstream electrode. This is likely due to the strong salt depletion in the downstream electrode, as presented in \cref{fig:ElectrodesOrder}a. Further, we see that for both configurations, by far the strongest electric field (potential gradient) occurs just downstream of the separator. As expected based on the results of \cref{fig:ElectrodesOrder}a, $c_{\mathrm{mA,\ce{B-}}}$ for the cat-an configuration is approximately zero across the cell, as the pH is too low for significant boric acid deprotonation. However, for the an-cat configuration we observe low values along most of the upstream electrode (anode), followed by a strong increase near the anode/separator interface, reaching a maximum value of 0.21 mM at the separator as boric acid is deprotonated in that region, followed by a decrease and lower values of approximately 0.04 mM in the cathode. Thus, due to strong boric acid deprotonation within the anode, the an-cat configuration is highly promising for effective boron electrosorption.

We argue that the counter-intuitive findings of these predictions are due to the complicated interplay between salt depletion, \ce{H+} and \ce{OH-} electrosorption and electromigration, and water splitting, see \cref{eq:ChemEquilWater}. For example, for the cat-an configuration, the low salt concentration $\left(c_{\mathrm{mA,NaCl}}<0.2\ \mathrm{mM}\right)$ in the anode leads to enhanced \ce{OH-} electrosorption into the micropores and can cause increases in local \ce{H+} production due to water splitting. The strong electric field in the upstream-end of the anode can increase \ce{H+} electromigration into the cathode, reducing pH in the cathode as seen in \cref{fig:ElectrodesOrder}a. Detailed study of such mechanisms are beyond the scope of the current work, but are important to unravel in future works which seek to optimize electrosorption of amphoteric species, such as boron, by CDI.

To explore the dynamics of our model CDI cell, \cref{fig:ElectrodesOrder}c shows the predicted pH at the anode/separator interface (left y-axis, black lines) and the cumulative electrosorbed boron by the cell, $\Gamma_{\mathrm{B}}$, defined in \cref{eq:GammaB} (right y-axis, blue lines), as functions of dimensionless time for the cat-an (solid lines) and an-cat (dashed lines) configurations. The pH at the anode/separator interface is a crucial metric, which can be used to rapidly assess whether the cell will be effective towards boric acid dissociation and boron ion electrosorption at the anode. For both configurations, we observe rapid development of the pH values at the anode/separator interface, within about $t/\tau_{\mathrm{D}}$ of 0.4, to near steady values. Thus, for essentially the entire charging process, the an-cat configuration is predicted to be capable of electrosorbing boron, as the solution at the anode/separator interface is alkaline. For the cat-an configuration, as expected the predicted boron electrosorption is negligible at all times during charging, while for the an-cat configuration, it is monotonically increasing for $t/\tau_{\mathrm{D}}\ge$ 0.5, reaching a value of $\SI{0.62}{\micro\mol/\gram}$ at $t/\tau_{\mathrm{D}}$=3. 

To validate the model predictions, we compare our theoretical findings with experimental data (see Materials and Methods for details). In \cref{fig:ElectrodesOrder}d we plot theoretical (lines) and experimental (markers) values of $\Gamma_{\mathrm{B}}$, as a function of Pe. We include predictions and measurements for boron removal in both cat-an (solid line, triangles) and an-cat (dashed line, circles) configurations. The theoretical predictions in \cref{fig:ElectrodesOrder}d show effective boron removal by the cell for the an-cat configuration, where no removal is expected for the cat-an configuration. While comparing theory and experimental results in \cref{fig:ElectrodesOrder}d, we observe a good agreement for the an-cat configuration for low and moderate flow velocities $\left(3\le \rm{Pe}\le 5\right)$, measuring a boron removal of around $\SI{0.50}{\micro\mol/\gram}$, although for high values $\left(\rm{Pe}=7\right)$ we obtain negative values of $\Gamma_{\mathrm{B}}$, indicating boron desorption. As predicted by the model, for all conditions tested, no significant boron removal was achieved in the cat-an configuration. Thus, the counter-intuitive prediction that an-cat is the preferred electrode order was also confirmed experimentally.

\subsection{Optimum charging voltage for boron removal}

\begin{figure}[!hbtp]
	\centering
	\includegraphics[width=0.8\textwidth]{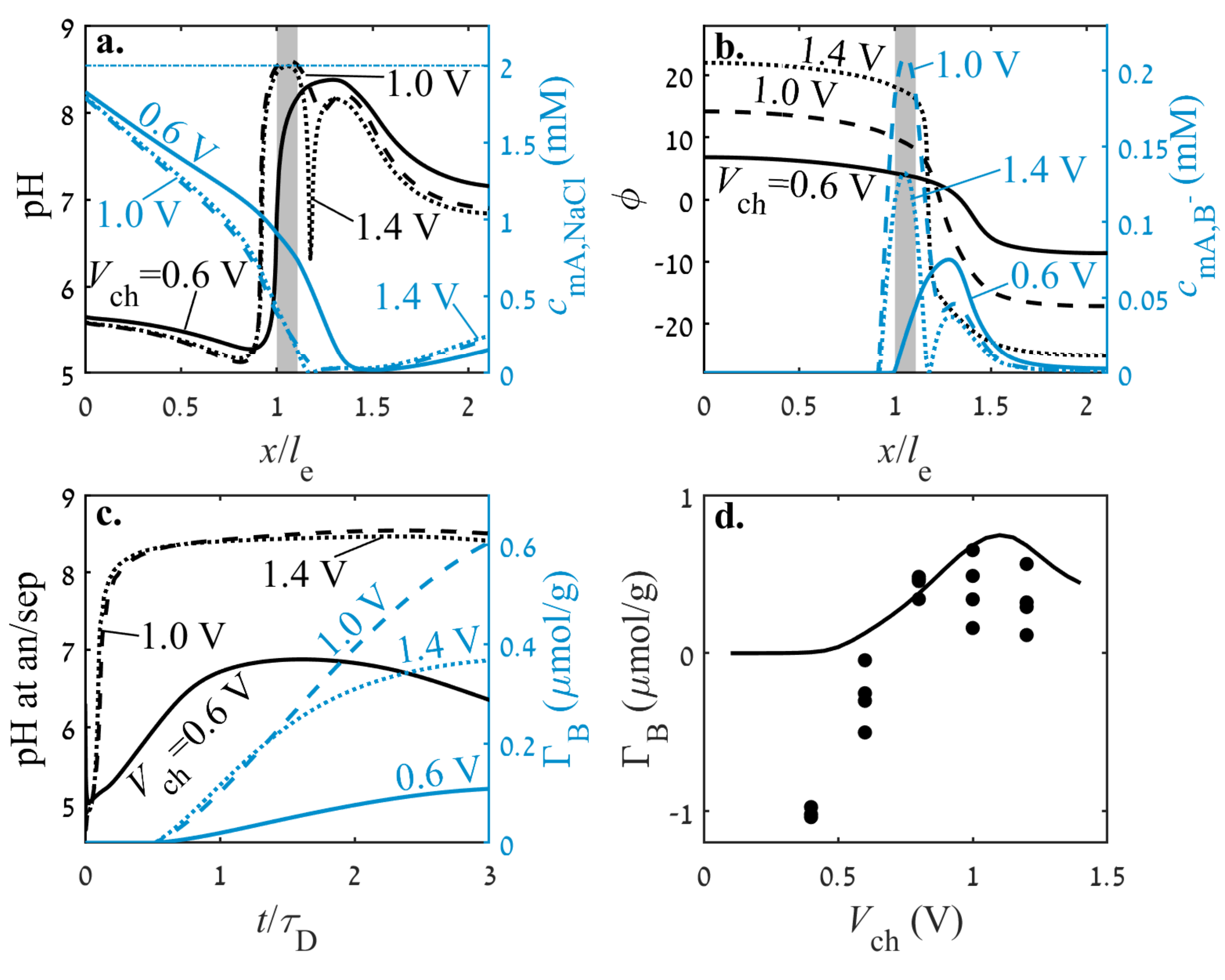}
	\caption{Predictions and experimental results for a FTE CDI cell with $c_{\mathrm{F}}$=2 mM, $c_{\mathrm{F,B}}$=0.37 mM and Pe=3. Predictions in panels a.-c. are for $V_{\mathrm{ch}}=0.6\ V$ (solid lines), 1.0 V (dashed lines) and 1.4 V (dotted lines). a. Predicted pH (left y-axis, black lines) and macropore salt concentration (right y-axis, blue lines) profiles. b. Predicted scaled potential (left y-axis, black lines) and \ce{B-} macropore concentration (right y-axis, blue lines) profiles. The profiles in panels a and b are for $t/\tau_D=2$, gray rectangles represent the separator, and the horizontal dash-dotted blue line in panel a represents feed salt concentration. c. Predicted pH at the anode/separator interface (left -axis, black lines) and electrosorbed boron (right y-axis, blue lines), both as a function of dimensionless time. d. Predicted (lines) and measured (markers) boron electrosorption for an effluent collection time of 10 min versus $V_{\mathrm{ch}}$.}
	\label{fig:Vcell}
\end{figure}

We further investigate the effects of charging voltage on pH dynamics and boron removal, now focusing on a cell with an-cat electrode order. Naively, we would expect to observe monotonic increase of boron removal with the charging voltage, as previous experimental observations show that pH perturbations in the anode and cathode become more extreme with increased cell voltage.\cite{Holubowitch2017,Landon2021} In \cref{fig:Vcell}a-c we show the predicted salt and charge dynamics of a cell operated with $c_{\mathrm{F}}$=2 mM, Pe=3 and charging voltages of 0.6 V (solid lines), 1.0 V (dashed lines) and 1.4 V (dotted lines), see \cref{tab:general,tab:electrodes} for other model parameters used. \cref{fig:Vcell}a shows the profiles of pH (left y-axis, black lines) and macropore salt concentration (right y-axis, blue lines) at $t/\tau_{\mathrm{D}}=2$. The pH profiles for all investigated charging voltages bear significant similarities: acidic values along most of the anode, followed by a strong rise near the anode/separator interface to alkaline values, and maintaining high pH across the separator and much of the cathode. For relatively high $V_{\mathrm{ch}}$, we observe a distinct pH local minima in the cathode near to the separator, with a minima of 8.0 for $V_{\mathrm{ch}}$=1.0 V, and much lower value of 6.3 for $V_{\mathrm{ch}}$=1.4 V. The salt concentration profiles for all the analyzed charging voltages are similar to the profile presented in \cref{fig:ElectrodesOrder}a for the an-cat configuration.

\cref{fig:Vcell}b presents the profiles of $\phi$ (left y-axis, black lines) and macropore \ce{B-} concentration (right y-axis, blue lines) at $t/\tau_{\mathrm{D}}=2$. As with \cref{fig:ElectrodesOrder}b, the potential profiles are similarly asymmetric about the cell midline, and the electric field is highest in the cathode near the separator. The location of the strongest electric field coincides with sharp local minima in pH as seen in \cref{fig:Vcell}a for the case of $V_{\mathrm{ch}}$=1.4 V. For all cell voltages analyzed, the macropore \ce{B-} concentration is near zero along most of the anode, followed by a sharp increase near the anode/separator interface. For $V_{\mathrm{ch}}$=0.6 V, the maximum of 0.075 mM occurs in the cathode. However, the maximum values of 0.21 mM for $V_{\mathrm{ch}}$=1.0 V and 0.13 mM for $V_{\mathrm{ch}}$=1.4 V are developed in the separator. Counter-intuitively, these results suggest that boron electrosorption will be less effective at $V_{\mathrm{ch}}$=1.4 V than $V_{\mathrm{ch}}$=1.0 V due to the slightly higher pH values developed for the latter case. 

To probe the effect of $V_{\mathrm{ch}}$ on the cell dynamics,  in \cref{fig:Vcell}c we plot predicted pH at the anode/separator interface (left y-axis, black lines) and cumulative electrosorbed boron (right y-axis, blue lines), both as a function of dimensionless time for charging voltages of 0.6 V (solid lines), 1.0 V (dashed lines) and 1.4 V (dotted lines). For $V_{\mathrm{ch}}$=0.6 V we observe relatively low pH values at this interface for the entire charging step, never surpassing 7. However, for $V_{\mathrm{ch}}$=1.0 V and 1.4 V we can see a fast increase in pH at at early times, $t/\tau_D<0.4$, followed by approximately steady pH at this interface of about 8.5 for $V_{\mathrm{ch}}$=1.4 V and slightly higher value of 8.6 for 1.0 V. The cumulative boron electrosorption for all the cases is monotonically increasing, but at $V_{\mathrm{ch}}$=0.6 V the amount of electrosorbed boron is the lowest, $\Gamma_{\mathrm{B}}$=$\SI{0.11}{\micro\mol/\gram}$ at $t/\tau_{\mathrm{D}}$=3, followed by $\SI{0.37}{\micro\mol/\gram}$ for $V_{\mathrm{ch}}$=1.4 V, and $\SI{0.60}{\micro\mol/\gram}$ at $V_{\mathrm{ch}}$=1.0 V.

Next, we compare these theoretical predictions to experimental measurements, see Materials and Methods for more details about the experimental setup. In \cref{fig:Vcell}d we compare between theoretical (lines) and experimental (circles) values of $\Gamma_{\mathrm{B}}$ as a function of $V_{\mathrm{ch}}$. The theoretical predictions in \cref{fig:Vcell}d show an optimum charging voltage for boron removal of approximately 1.1 V. Good quantitative agreements between theory and experiments are obtained for high charging voltages ($V_{\mathrm{ch}}\ge$0.8 V), however at lower voltages the model predicts no boron storage, and the experiments show significant boron desorption. Overall, experiments and theory agree and show that increasing cell voltage does not necessarily lead to improved boron removal, and that an optimum cell voltage is achieved at around 1 V.

\subsection{Effects of feed salt concentration and velocity on local pH}
\begin{figure}[t]
    \centering
    \includegraphics[width=0.8\textwidth]{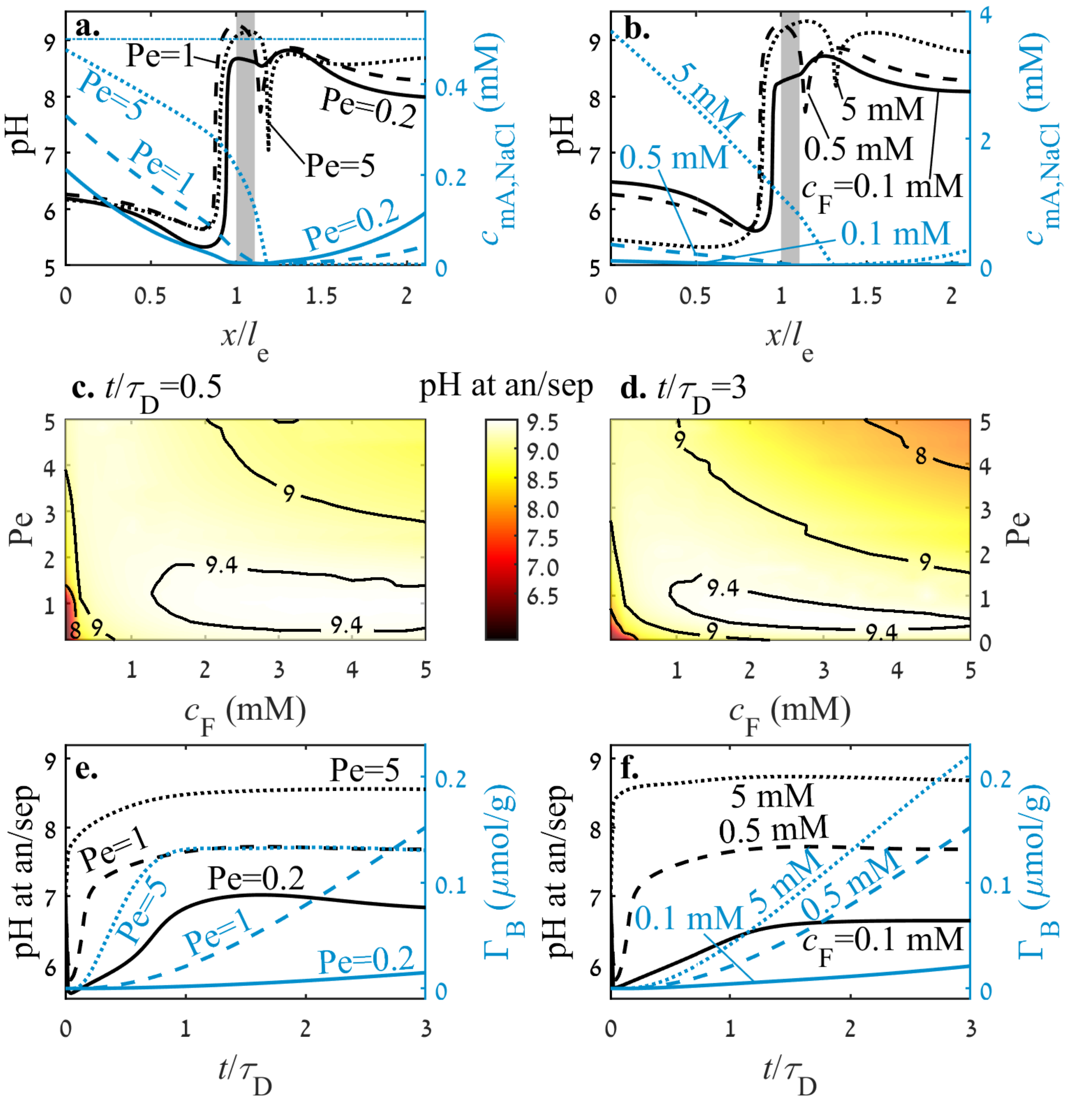}
    \caption{Predictions for a FTE CDI cell with with NaCl-containing feedwater (a.-d.), or NaCl and boron-containing feedwater (e.-f.), with $V_{\mathrm{ch}}$=1.0 V. a. and b. Predicted pH (left y-axes, black lines) and macropore salt concentration (right y-axes, blue lines) profiles at $t/\tau_D=3$, where the gray rectangles represent the separator, and the horizontal dash-dotted blue line in a. represents the feed salt concentration. a. and e. Results for ${\rm Pe}=0.2$ (solid lines), 1 (dashed lines), and 5 (dotted lines), all with $c_{\mathrm{F}}$=0.5 mM. c. and f. Results for $c_{\mathrm{F}}$=0.1 mM (solid lines), 0.5 mM (dashed lines), and 5 mM (dotted lines), all with Pe=1. c. and d. Predicted pH at the anode/separator interface as a function of $c_{\mathrm{F}}$ and Pe at $t/\tau_{\mathrm{D}}=0.5$ (c.) and $t/\tau_{\mathrm{D}}=3$ (d.).}
    \label{fig:NoB_PeCfeed}
\end{figure}

Having established that our theory captures key trends in boron electrosorption, we now investigate theoretically the effects of the flow velocity and feed salt concentration on pH dynamics and boron electrosorption. In \cref{fig:NoB_PeCfeed}a-d we present the analysis of a cell operated with solely salt in the feedwater and no amphoteric species, where in \cref{fig:NoB_PeCfeed}e-f we include boric acid in the feed. Model parameters used are tabulated in \cref{tab:general,tab:electrodes}. \cref{fig:NoB_PeCfeed}a presents the effect of the flow velocity, quantified by Pe, on the pH in the cell, plotting the pH (left y-axis, black lines) and macropore salt concentration (right y-axis, blue lines) profiles at $t/\tau_D=3$, for Pe=0.2 (solid lines), 1 (dashed lines) and 5 (dotted lines), all with $c_{\mathrm{F}}$=0.5 mM and $V_{\mathrm{ch}}$=1.0 V. For the case of low flow velocity, ${\rm Pe}=0.2$, in \cref{fig:NoB_PeCfeed}a, we observe very low salt concentrations across the whole cell, with near zero concentration in the separator. For moderate and high flow velocities, Pe=1 and 5, we observe a decreasing salt concentration across the anode, followed by very low concentration throughout the cathode. For all Pe tested, the pH profiles of the three analyzed cases show low pH values across most of the anode of approximately 6.0, and high pH values of approximately 8.5 across the separator and cathode. 

\cref{fig:NoB_PeCfeed}b plots the pH (left y-axis, black lines) and macropore salt concentration (right y-axis, blue lines) profiles at $t/\tau_D=3$, for $c_{\mathrm{F}}=0.1\ \mathrm{mM}$ (solid lines), 0.5 mM (dashed lines), and 5 mM (dotted lines), all with ${\rm Pe}=1$ and $V_{\mathrm{ch}}$=1.0 V. For all salt concentrations, the pH profiles in \cref{fig:NoB_PeCfeed}b are similar: low pH values in the anode, followed by an increase near the anode/separator interface, and high pH values across the separator and the cathode. The pH at the anode/separator interface for $c_{\mathrm{F}}=0.1\ \mathrm{mM}$ is 8.3, 9.1 for $5\ \mathrm{mM}$ and 9.3 for $c_{\mathrm{F}}=2\ \mathrm{mM}$. To conveniently visualize that the effect of flow velocity and feed salt concentration on the important metric of pH at the anode/separator interface, in \cref{fig:NoB_PeCfeed}c and d we present colormaps and contour plots at $t/\tau_\mathrm{D}=0.5$ (\cref{fig:NoB_PeCfeed}c) and $t/\tau_\mathrm{D}=3$ (\cref{fig:NoB_PeCfeed}d). For early times the pH at the anode/separator interface is above 9 for most of the investigated parameter range, while for later times, the pH decreases somewhat at high values of $c_\mathrm{F}$ and Pe. However, over almost the entire parameter set investigated here, pH at the anode/separator interface is high enough to be suitable for significant boric acid dissociation and boron electrosorption at the anode.

In \cref{fig:NoB_PeCfeed}e and f we extend the analysis for a feedwater with both NaCl and boron. \cref{fig:NoB_PeCfeed}e and f plot the pH at the anode/separator interface (left y-axis, black lines) and $\Gamma_{\mathrm{B}}$ (right y-axis, blue lines), both as a function of dimensionless time. \cref{fig:NoB_PeCfeed}e shows results for Pe=0.2 (solid lines), 1 (dashed lines) and 5 (dotted lines), all with $c_{\mathrm{F}}$=0.5 mM, while \cref{fig:NoB_PeCfeed}f shows results for $c_{\mathrm{F}}=0.1$ mM (solid lines), 0.5 mM (dashed lines), and 5 mM (dotted lines), all with Pe=1.

The pH values for Pe=0.2 in \cref{fig:NoB_PeCfeed}e are relatively low along the whole process, followed by the values for Pe=1, where for Pe=5 a strong pH increase is observed at early times reaching values above 8.5, followed by a steady and moderate increase. For Pe=0.2 and 1 we observe monotonic increase of $\Gamma_{\mathrm{B}}$ along the investigated time range, while for Pe=5 we observe strong boron removal at $t/\tau_{\mathrm{D}}\le$0.9, followed by a steady value. The pH values at the anode in \cref{fig:NoB_PeCfeed}e are similar to the values presented for a cell operated without boron, where the presence of boron in the system restrains the pH rise. The boron removal values presented in \cref{fig:NoB_PeCfeed}e are logical, as the boron adsorption increases with pH. However, the lack of boron removal for Pe=5 at $t/\tau_{\mathrm{D}}>$0.9 may be due to the relatively high \ce{Cl-} concentration near the anode/separator interface, which competes with boron to be stored in anode micropores.

The pH values for $c_{\mathrm{F}}$=0.1 mM presented in \cref{fig:NoB_PeCfeed}f are low at all times during charging, significantly higher pH seen for $c_{\mathrm{F}}$=0.5 mM. For $c_{\mathrm{F}}$=5 mM the pH reaches about 8.7 for $t/\tau_{\mathrm{D}}\ge$ 0.2. Thus, boron removal is very low for $c_{\mathrm{F}}$=0.1 mM, while for $c_{\mathrm{F}}$ of 0.5 mM and 5 mM higher values are predicted, reaching $\SI{0.22}{\micro\mol/\gram}$ at $t/\tau_{\mathrm{D}}$=3 for the latter case. 

\section{Conclusions}
The removal of amphoteric species from water is important, both for water purification and for resource recovery, but is often challenging to accomplish with conventional membrane-based technologies. We here propose and establish the theory for removal of amphoteric species by membraneless capacitive deionizaton (CDI) cells, which generate large internal pH gradients during cell charging. Our proposed theoretical framework extends traditional CDI theory by including both pH dynamics and amphoteric species in the feedwater that can protonate or deprotonate depending on local pH. Both our model and validating experiments showed important and counter-intuitive design rules for removal of boron, our example amphoteric species. Despite conventional wisdom in CDI, we found that the anode should be placed upstream to achieve effective boron removal, and that increasing cell voltage does not necessarily lead to improved boron removal. Although generally good agreement between experiments and theory were attained for key trends, quantitative differences between theory and experiments are hypothesized to be due to Faradaic side-reactions which were not included in the model presented here. In the future, the framework here can be extended to include Faradaic side-reactions to enhance its quantitative predictive capabilities, and applied to any number of amphoteric species.

\section{Materials and Methods}
\subsection{Experimental apparatus}
Activated carbon cloth (ACC-5092-15, Kynol GmbH, Germany) with $\SI{600}{\micro\meter}$ thickness,  $\SI{0.53}{\milli\liter/\gram}$ specific micropore volume, and approximately $\SI{1400}{\meter\squared/\gram}$ surface area (latter two values from Ref. \cite{Guyes2021}) was used as the electrode material. This material was characterized in a number of previous CDI studies.\cite{Bouhadana2011,Kim2017,Guyes2019,Uwayid2020,Guyes2021} The carbon was cut into electrode squares with a cross-sectional area of $\SI{6.25}{\centi\meter\squared}$. Electrodes were rinsed with deionized water, dried for 3 hr at $\SI{80}{\celsius}$, then weighed immediately.

The FTE CDI cell was described and illustrated in Ref. \cite{Guyes2021}. Briefly, the cell consists of two electrodes electronically isolated by a separator (Omnipore JHWP PTFE membrane, Merck, $\SI{65}{\micro\meter}$ thickness) cut to a size of  $\SI{2.7}{\centi\meter}$ $\times$ $\SI{2.7}{\centi\meter}$. Graphite current collectors contact the posterior side of each electrode, and a grid of holes was milled into each collector to allow water passage. The upstream reservoir is located just ahead of the first current collector, and likewise the downstream reservoir is located just behind the second collector. The cell is enclosed on both sides by milled PVDF blocks, which contain ports for fluid flow and air evacuation. Compressible ePTFE gaskets seal the cell.

Feedwater composed of 0.37 mM ($\SI{4}{\milli\gram/\liter}$) boric acid (Bio-Lab, Israel) and 2 mM NaCl ($>$99.5\%, SDFCL, India) was prepared with $\SI{18.2}{\mega\ohm}$ deionized water (Synergy, Merck KGaA, Germany). Prior to each CDI experiment, the feedwater was purged of dissolved oxygen by nitrogen gas bubbling for 20 minutes in a 0.5 L glass feed reservoir under stirring. A peristaltic pump (Masterflex 07551-30, Cole-Parmer, USA) then transported the feedwater through the conductivity sensor (Tracedec 390-50, Innovative Sensor Technologies GmbH, Austria to measure its initial conductivity, and a pH electrode (iAquatrode Plus Pt1000, Metrohm AG, Switzerland) was dipped into the feed reservoir to measure the initial feedwater pH.

In each CDI experiment, the peristaltic pump supplied feedwater to the CDI cell. The electrodes were initially discharged at 0 V with a voltage source (2400 Source Meter, Keithley Instruments, USA) while feedwater flowed through the cell until the current was negligible. The cell was then charged for 12 min at a given voltage and flow rate and discharged for 30 min at 0 V for three consecutive charge-discharge cycles. The long discharge step allowed the cell to re-equilibrate with the incoming feedwater before the subsequent charge step. Effluent conductivity was measured for the duration of the experiment. During the charge step of the 3rd (limit) cycle, the effluent solution was collected continuously for 10 min beginning from the moment when the dynamic effluent conductivity decreased below the feed conductivity value. Feedwater samples were also collected for analysis.

The boron concentrations in the collected samples were measured with the Azomethine-H method of L{\'{o}}pez et  al.\cite{Lopez1993} Absorbance was measured at 414 nm against a blank reference solution with an Evolution 300 spectrophotometer (Thermo Fisher Scientific, USA). Boron concentration was interpolated from a calibration curve constructed from the absorbances of 1-5 $\SI{}{\milli\gram/\liter}$ standard solutions.

\subsection{Theory}
\begin{figure}[!htbp]
	\centering
	\includegraphics[width=0.9\textwidth]{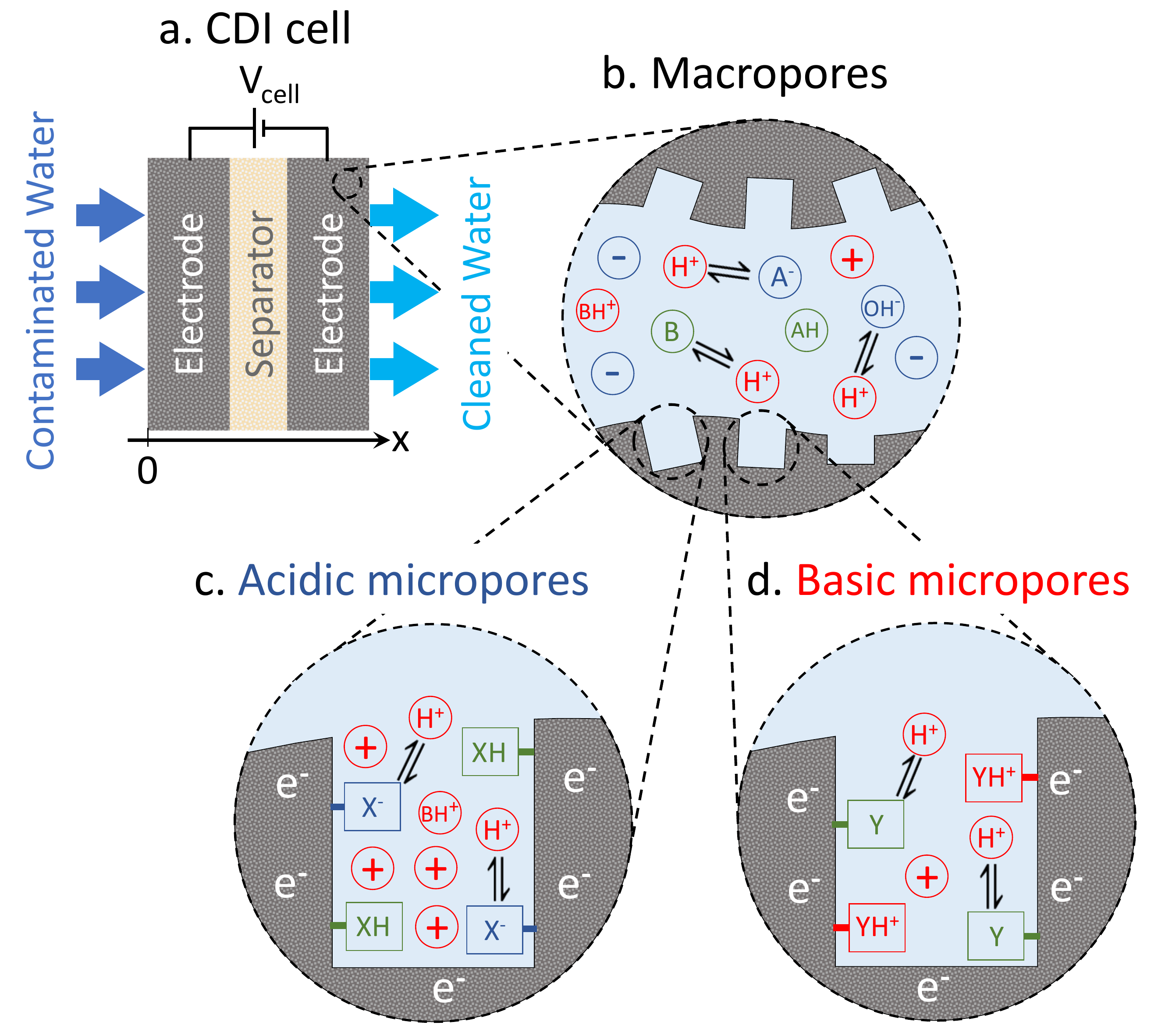}
	\caption{Schematic overview of a CDI cell for the removal of amphoteric species. a. The FTE CDI cell consists of two electrodes separated by a separator, through which feedwater, which contains inert and amphoteric ions, flows while the electrodes electrosorb the ions. b. In the macropores, which serve as transport pathways, the local pH across the cell is evaluated, as well as the pH-dependent chemical equilibria between the amphoteric ions, and the \ce{H+} and \ce{OH-} ions. c. and d. Micropores for ion adsorption into overlapping EDLs, filled by electrosorbed ions, and chemical surface groups with pH-dependent chemical charge. Several regions are considered, including acidic (c.) and basic (d.) regions.}
	\label{fig:TheorySchematics}
\end{figure}

To describe the transport and removal of amphoteric species in an electrochemical cell employing porous carbon electrodes, we present a novel theoretical framework that describes the joint processes: I) ion transport due to advection, diffusion, and migration, II) ongoing association and dissociation equilibrium reactions of water and amphoteric species, and III) ion adsorption in electrical double layers (EDLs), including the presence of  pH-dependent chemical surface groups. In the porous electrodes, we distinguish two different types of pores: macropores, which serve as transport pathways, see \cref{fig:TheorySchematics}b, and micropores, in which ions are electrosorbed into overlapping EDLs, see \cref{fig:TheorySchematics}c and d. These definitions are based on pore function, whereas the conventionally used definitions by the IUPAC are based on pore size.\cite{Porada2013a}

To describe ion electrosorption in the porous carbon electrodes, we use the multi-equilibria amphoteric modified Donnan (ME-amph-mD) model.\cite{Legrand2020} This model is an extended version of the amphoteric modified Donnan (amph-mD) model, which has been extensively validated and shown to describe experimental data well.\cite{Mubita2019,Dykstra2017} The amph-mD model considers the presence of a fixed number of chemical groups in the micropores, but omits the pH-dependency of these groups. In contrast, the ME-amph-mD model includes this pH-dependency, based on the models presented by Hemmatifar et al. and Oyarzun et al.\cite{Hemmatifar2017,Oyarzun2018a}

In the ME-amph-mD model, as well as in the amph-mD model, we consider several micropore regions: $m$ acidic regions and $n$ basic regions. In these micropore regions, surface groups with a pH-dependent charge are present. In the present work, we only consider monovalent chemical groups, reacting only with \ce{H+} or \ce{OH-} ions, see \cref{fig:TheorySchematics}c and d. We follow Persat et al.\cite{Persat2009a} and Hemmatifar et al.,\cite{Hemmatifar2017} and consider a local chemical equilibrium between the protonated and deprotonated surface groups, which dependents on an equilibrium constant, $K_a$, and the local pH. In the acidic regions, negatively charged chemical surface groups are present, marked as $\ce{X-}$ in \cref{fig:TheorySchematics}c, while in the basic regions, positively charged groups are present, marked as $\ce{YH+}$ in \cref{fig:TheorySchematics}d. All these groups can be in a protonated or deprotonated state, dependent on the local pH in the respective micropore region. The state of the groups is found by the following chemical equilibrium reactions

\begin{align}
    \ce{X_{\it i}H \rightleftharpoons X_{\it i}- + H+},& K_{\ce{X_{\it i}H}}=\frac{\left[\ce{X_{\it i}-}\right]\cdot\left[\ce{H+}\right]}{\left[\ce{X_{\it i}H}\right]}
    \label{eq:AcidReac}\\
    \ce{Y_{\it j}H+ \rightleftharpoons Y_{\it j} + H+},& K_{\ce{Y_{\it j}H+}}=\frac{\left[\ce{Y_{\it j}}\right]\cdot \left[\ce{H+}\right]}{\left[\ce{Y_{\it j}H+}\right]}
    \label{eq:BaseReac}
\end{align}
where $K_{\ce{X_iH}}$ is the equilibrium constant of the $i$-th acidic region, and $K_{\ce{Y_jH+}}$ is the equilibrium constant of the $j$-th basic group.\footnote{Please note that we use the notation $[i]$ to denote an ionic concentration, $c_{i}$, and that we use $[i]$ and $c_{i}$ interchangeably.} The chemical surface charge in these regions is pH-dependent, i.e., only a part of the total concentration of chemical surface groups is charged, and the rest are not charged. The acidic groups are negatively charged in their deprotonated state, and the basic surface groups are positively charged in their protonated state. The chemical surface charge is thus given by

\begin{align}
    \sigma_{\mathrm{chem,X_{\it i}}} &= -\frac{c_{\mathrm{chem,X_{\it i},t}}}{1+c_{\mathrm{mi,\ce{H+},X_{\it i}}}/K_{\ce{X_{\it i}H}}}
    \label{eq:ChargeAcidReac}\\
    \sigma_{\mathrm{chem,Y_{\it j}}} &= \frac{c_{\mathrm{chem,Y_{\it j},t}}}{1+K_{\ce{Y_{\it j}H+}}/c_{\mathrm{mi,\ce{H+},Y_{\it j}}}}
    \label{eq:ChargeBaseReac}
\end{align}
where subscript $\ce{X_i}$ refers to the $i$-th acidic region, and $\ce{Y_j}$ to the $j$-th basic region, and where $\sigma_{\mathrm{chem},R}$ is the chemical surface charge, $c_{\mathrm{chem},R,\mathrm{t}}$ the total concentration of chemical surface groups, and $c_{\mathrm{mi,\ce{H+}},R}$ is the \ce{H+} concentration, all in the micropore in region $R$.

The ion concentration of the $i$-th ion in the micropores in region $R$ is given by considering local thermodynamic equilibrium
\begin{equation}
    c_{\mathrm{mi},i,R}=c_{\mathrm{mA},i} \cdot \mathrm{exp} \left(-z_i \Delta \phi_{\mathrm{D},R}\right)
    \label{eq:amphD}
\end{equation}
where $c_{\mathrm{mi},i,R}$ is the concentration in region $R$, $c_{\mathrm{mA},i}$ is the concentration in the macropores, $z_i$ is the ion valence, and $\Delta\phi_{\mathrm{D},R}$ is the Donnan potential in region $R$. The Donnan potential, as all other potentials presented below, can be multiplied by the thermal voltage, given by $V_\mathrm{T}\equiv k_\mathrm{B}T/e$, to arrive at a dimensional voltage, where ${k_\mathrm{B}}$ is the Boltzmann constant, $T$ is the absolute temperature and $e$ is the elementary charge.

We define the total micropore concentration for each species by summing over all micropore regions the concentration of free ions in the micropores and the concentration of ions chemically bound to the surface groups. The total concentration is calculated by
\begin{equation}
    c_{\mathrm{mi,tot},i}=\sum_R \alpha_{\mathrm{mi},R}\left( c_{\mathrm{mi},i,R}+\beta_{\mathrm{mi},i,R}\cdot c_{\mathrm{chem},R,\mathrm{t}}\right)
    \label{eq:Cmi}
\end{equation}
where $\alpha_{\mathrm{mi},R}$ is the fraction of the total micropore volume, that is in region $R$, and $\beta_{\mathrm{mi},i,R}$ is the ratio between the concentration of $i$-th ion bound to the chemical groups, and the total concentration of chemical groups, both in region $R$. Here, we account for chemical groups reacting only with \ce{H+} or \ce{OH-} ions, so $\beta_{\mathrm{mi},i,R}=0$ holds for all species, except for \ce{H^+} and \ce{OH^-}. Moreover, under the convention of considering only acid dissociation constants, see \cref{eq:AcidReac,eq:BaseReac}, we note that $\beta_{\mathrm{mi},\ce{OH-},R}=0$, and
\begin{equation}
    \beta_{\mathrm{mi},\ce{H+},R}=\frac{1}{1+K_R/c_{\mathrm{mi,\ce{H+}},R}}
    \label{eq:betaExpression}
\end{equation}
where $K_R$ is the reaction constant of the chemical group in region $R$. Next, by combining \cref{eq:betaExpression} with \cref{eq:AcidReac,eq:BaseReac} the following expressions hold
\begin{equation}
    \sigma_{\mathrm{chem,X_i}} = -\left(1-\beta_{\mathrm{mi},\ce{H+}\mathrm{X_i}}\right)\cdot c_{\mathrm{chem,X_i,t}}
    \label{eq:SigmaBeta_Acid}
\end{equation}
\begin{equation}
    \sigma_{\mathrm{chem,Y_j}} = \beta_{\mathrm{mi},\ce{H+}\mathrm{Y_j}}\cdot c_{\mathrm{chem,Y_j,t}}.
    \label{eq:SigmaBeta_Base}
\end{equation}

For each micropore region, we can calculate the ionic charge, given by
\begin{equation}
    \sigma_{\mathrm{ionic},R}=\sum_{i} z_i\cdot c_{\mathrm{mi},i,R} .
\end{equation}
Moreover, for each micropore region, a charge balance holds, which states that the sum of electronic charge, ionic charge and chemical charge is equal to zero
\begin{equation}
    \sigma_{\mathrm{elec},R}+\sigma_{\mathrm{chem},R}+\sigma_{\mathrm{ionic},R}=0
\end{equation}
where $\sigma_{\mathrm{elec},R}$ is the electronic charge in region $R$. The electronic charge is related to the Stern capacitance and the Stern potential by
\begin{equation}
    \sigma_{\mathrm{elec},R} \cdot F= V_{\mathrm{T}} \cdot C_{\mathrm{S},R} \Delta \phi_{\mathrm{S},R}
\end{equation}
where $F$ is the Faraday constant, $C_{\mathrm{S},R}$ is the Stern capacitance and $\Delta \phi_{\mathrm{S},R}$ is the Stern potential, both in region $R$.

For each electrode, and each micropore region, the summation of $\Delta \phi_{\mathrm{D},R}$ and $\Delta \phi_{\mathrm{S},R}$ is equal to the difference between the electrical potential in the electrode and the macropores. Therefore, we evaluate the following equation for each electrode
\begin{equation}
    \phi_{\mathrm{E},e}-\phi_{\mathrm{mA}} = \Delta \phi_{\mathrm{D},R} + \Delta \phi_{\mathrm{S},R}
\end{equation}
where $\phi_{\mathrm{mA}}$ is the dimensionless potential in the macropores, subscript ``E'' refers to the electrode region and subscript ``e'' refers to the electrode type, which is either the anode (an) or the cathode (cat). The values of $\phi_{\mathrm{E,an}}$ and $\phi_{\mathrm{E,cat}}$ are related to the cell voltage by the expression
\begin{equation}
     V_{\mathrm{T}}(\phi_{\mathrm{E,an}} - \phi_{\mathrm{E,cat}}) = V_\mathrm{cell}-I\cdot \rm{EER}
\end{equation}   
where $I$ is the electric current through the cell and EER refers to external electronic resistance \cite{Dykstra2016}.

To describe the transport of ions, we use the Nernst-Planck equation
\begin{equation}
    J_{i}=v_{\mathrm{w}}c_{\mathrm{mA},i} -D_{\mathrm{mA},i} \left(\frac{\partial c_{\mathrm{mA},i}}{\partial x} +z_i \cdot c_{\mathrm{mA},i} \frac{\partial \phi_{\mathrm{mA}}}{\partial x}\right)
\end{equation}
where $J_i$ is the superficial flux of ionic species $i$, $v_{\mathrm{w}}$ is the superficial velocity of the water phase, $D_{\mathrm{mA},i}$ is the effective ion diffusion coefficient in the macropores, and $x$ is the location in the cell, as shown in \cref{fig:TheorySchematics}a. The parameter $D_{\mathrm{mA},i}$ is given by $D_{\mathrm{mA},i}\equiv D_{\infty,i} \cdot p_{\mathrm{mA}}/\tau_{\mathrm{mA}}$, where $D_{\infty,i}$ is the diffusion coefficient in free solution, $p_{\mathrm{mA}}$ is the macropore porosity and $\tau_{\mathrm{mA}}$ is the macropore tortuosity.

Mass conservation holds in the electrodes. Therefore, we evaluate the following mass balance equation
\begin{equation}
    p_{\mathrm{mA}}\frac{\partial c_{\mathrm{mA},i}}{\partial t}+p_{\mathrm{mi}}\frac{\partial c_{\mathrm{mi,tot},i}}{\partial t}=-\frac{\partial J_{i}}{\partial x} + \Gamma_i
    \label{eq:massbal}
\end{equation}
where $p_\mathrm{mi}$ is the micropore porosity, and  $c_{\mathrm{mi,tot},i}$ is given by \cref{eq:Cmi}. Moreover, $\Gamma_i$ is the production rate of the $i$-th ion due to acid-base equilibrium reactions in the bulk. For inert ions, which do not participate in acid-base equilibria, such as \ce{Na+} and \ce{Cl-}, $\Gamma_i=0$. For amphoteric ions, the value of $\Gamma_i$ is almost always different from zero, and the value is often unknown. However, although $\Gamma_i$ is unknown, an explicit evaluation of $\Gamma_j$ is not required, as presented in the next paragraph.

Let us illustrate how $\Gamma_i$ disappears, following the approach described in Refs. \cite{Bercovici2009,DeLichtervelde2019,Biesheuvel2020} We derive a total mass balance equation for a group of species, marked by $G$, which includes all species that are in chemical equilibrium with each other. At each position in the system, the summation of the production rates, $\Gamma_i$, of all species in a group, $G$, is equal to zero, i.e., $\sum\limits_{i\in G}\Gamma_i=0$. Consequently, summing \cref{eq:massbal} over the individual species within group $G$, all $\Gamma _i$ terms cancel out, resulting in the following mass balance equation
\begin{equation}
    \sum_{i\in G}{\left(p_{\mathrm{mA}}\frac{\partial c_{\mathrm{mA},i}}{\partial t}+p_{\mathrm{mi}}\frac{\partial c_{\mathrm{mi,tot},i}}{\partial t}\right)}=-\sum_{i\in G}\frac{\partial J_{i}}{\partial x}.
    \label{eq:massbalGroup}
\end{equation} 
The approach presented by \cref{eq:massbalGroup} eliminates the $\Gamma_i$ terms from the system of equations that is solved. To solve the resulting set of equations, we assume the acid-base equilibrium reactions between ionic species within a group are infinitely fast, i.e., these species are in chemical equilibrium. Thus we evaluate, at each position in the system and at each moment, the following chemical equilibrium conditions for the group
\begin{equation}
    K_i=\frac{\left[\mathrm{P}_i\right]\cdot\left[\ce{H+}\right]}{\left[\mathrm{R}_i\right]}
    \label{eq:ChemEquilG}
\end{equation}
where $K_i$ is the equilibrium constant, $\mathrm{P}_i$ is the product, and $\mathrm{R}_i$ is the reactant, all of the $i$-th acid-base equilibrium reaction. Similarly, we consider the water dissociation equilibrium, \ce{H2O \rightleftharpoons H+ + OH-}, with the equilibrium constant 
\begin{equation}
    K_{\mathrm{w}} = \left[\ce{H+}\right] \cdot \left[\ce{OH-}\right].
    \label{eq:ChemEquilWater}
\end{equation}

Now, we consider electroneutrality at each position in the macropores and in the separator
\begin{equation}
    \sum_{i} z_{i} \cdot c_{\mathrm{mA},i} = 0.
    \label{eq:electroneutrality}
\end{equation} 
Next, we derive the charge balance by multiplying \cref{eq:massbal} by $z_i$, and summing the resulting equations over all ionic species, including \ce{H+} and \ce{OH-}
\begin{equation}
    p_{\mathrm{mi}} \frac{\partial}{\partial t} \sum_{i} z_{i} \cdot c_{\mathrm{mi,tot},i} = - \frac{\partial }{\partial x} \sum_{i} z_{i} J_{i}.
    \label{eq:chargebal}
\end{equation}
Only the micropores contribute to the left hand side of \cref{eq:chargebal}, as charge does not accumulate in the macropores, see \cref{eq:electroneutrality}. Furthermore, charge conservation in the equilibrium reactions eliminates the term associated with the reaction rates. Also, by expanding \cref{eq:chargebal}, and substituting \cref{eq:SigmaBeta_Acid,eq:SigmaBeta_Base} into it, the charge balance can be written as
\begin{equation}
    p_{\mathrm{mi}} \sum_R{\alpha_{\mathrm{mi},R}\bigg( \frac{\partial \sigma_{\mathrm{chem},R}}{\partial t} +\frac{\partial \sigma_{\mathrm{ionic},R}}{\partial t}\bigg)}=-\frac{\partial J_{\mathrm{ch}}}{\partial x}
    \label{eq:chargebalDet}
\end{equation}
where $J_{\mathrm{ch}}$ is the superficial charge flux, defined as $J_{\mathrm{ch}}\equiv \sum\limits_i{z_i\cdot J_i}$. In the separator, $J_{\mathrm{ch}}$ is related to $I$, by the expression $I=F\cdot A_\mathrm{c}\cdot J_{\mathrm{ch}}$, where $A_\mathrm{c}$ is the electrode cross section area.

Next, we describe ion transport across the separator, a thin layer of insulating porous material allowing the transport of water and ions. To do so, we use a mass balance equation for all the inert ions (\cref{eq:massbal}), a mass balance for all amphoteric groups (of the form of \cref{eq:massbalGroup}), the charge balance (\cref{eq:chargebalDet}), the electroneutrality condition (\cref{eq:electroneutrality}), and the chemical equilibrium equations (\cref{eq:ChemEquilG,eq:ChemEquilWater}). In the aforementioned equations, the subscript ``mA'' is replaced by ``sep'', referring to the separator, and $p_{\mathrm{mi}}=0$.  

On both separator/electrode interfaces, we consider continuity of flux for the inert ions, $\left. J_{\mathrm{sep},i} \right|_{\mathrm{sep/e}} =  \left. J_{\mathrm{e},i} \right|_{\mathrm{sep/e}} \label{eq:BCinelec}$, where the subscripts ``$\mathrm{sep}$'' and ``$\mathrm{e}$'' represent the separator side and electrode side, respectively. Similarly, we consider continuity of flux for the groups of amphoteric ions, $\sum_{i\in\ce{G}} \left. J_{\mathrm{sep},i} \right|_{\mathrm{sep/e}} =  \sum_{i\in\ce{G}} \left. J_{\mathrm{e},i} \right|_{\mathrm{sep/e}}$, and charge flux, $ \left. J_{\mathrm{sep,ch}} \right|_{\mathrm{sep/e}} =  \left. J_{\mathrm{e,ch}} \right|_{\mathrm{sep/e}}$. Moreover, we consider continuity of concentration, $\left. c_{\mathrm{sep},i} \right|_{\mathrm{sep/e}} =  \left. c_{\mathrm{mA},i} \right|_{\mathrm{sep/e}}$, and potential , $\left. \phi_{\mathrm{sep},i} \right|_{\mathrm{sep/e}} =  \left. \phi_{\mathrm{mA},i} \right|_{\mathrm{sep/e}}$, in both separator/electrode interfaces.

Before solving the set of equations described in preceding paragraphs, we must specify the boundary and initial conditions. To specify the proper boundary conditions, we first estimate the P{\'{e}}clet number. For cases where ${\rm Pe}\gg 1$, the approximation described in Ref. \cite{Guyes2017Model} can be used. For other cases, the salt and charge dynamics in the upstream and downstream reservoirs should be described by solving \cref{eq:massbal,eq:massbalGroup,eq:electroneutrality,eq:chargebalDet}, with $p_{\mathrm{mi}}=0$. In this case, $c_i=c_{\mathrm{F},i}$ and $J_{\mathrm{ch}}=0$ conditions are applied at the entrance to the upstream reservoir, where $c_{\mathrm{F},i}$ is the feed concentration of the $i$-th ion. Moreover, $J_i=v_{\mathrm{w}}c_i$ and $J_{\mathrm{ch}}=0$ conditions are applied at the downstream reservoir's exit.

To find the initial conditions, we assume the system is in equilibrium before the charging step begins. Therefore, all bulk concentrations are equal to the feed values, the potential in the macropores and separator is uniform across the whole system, $\phi_{\mathrm{mA}} = \phi_{\mathrm{sp}} = \phi_{\mathrm{mA},0}$, and the cell voltage is equal to the equilibrium value before charging. To find $\phi_{\mathrm{mA},0}$, the charge balance equation in the system must be solved, while remembering that $\sigma_{\mathrm{elec,R}}$ is a function of the ions concentrations and potential in the macropores
\begin{equation}
    l_{\mathrm{an}} \left. \left(p_{\mathrm{mi}} \sum_R{\alpha_{\mathrm{mi},R} \sigma_{\mathrm{elec},R}}\right)\right|_{\mathrm{an}}=l_{\mathrm{cat}} \left. \left(p_{\mathrm{mi}} \sum_R{\alpha_{\mathrm{mi},R} \sigma_{\mathrm{elec},R}}\right)\right|_{\mathrm{cat}}
\end{equation}
where $l_{\mathrm{an}}$ is the anode thickness and $l_{\mathrm{cat}}$ is the cathode thickness.

\subsection{Theoretical description of the experiment}
We add several extensions to the theoretical framework presented in the previous subsection for precise description of the experimental apparatus used in this work. First, we follow the method described by Guyes et al.\cite{Guyes2017Model} and account for the mixing of the solution before reaching the conductivity sensor placed downstream the cell
\begin{equation}
    V_{\mathrm{mix}}\frac{\partial c_{\mathrm{cs},i}}{\partial t} = A_{\mathrm{c}}v_{\mathrm{w}}\left(c_{\mathrm{mA},i,\mathrm{eff}}-c_{\mathrm{cs},i}\right)
\end{equation}
where $V_{\mathrm{mix}}$ is the mixing volume, $c_{\mathrm{cs},i}$ is the concentration at the conductivity sensor and $c_{\mathrm{mA},i,\mathrm{eff}}$ is the effluent concentration, both of the $i$-th species. Moreover, collection of treated solution begins only when the conductivity value at the sensor reaches values below the feed conductivity, imitating the experimental approach. Last, we quantify the electrosorbed boron, $\Gamma_{\mathrm{B}}$, using the following definition
\begin{equation}
    \Gamma_{\mathrm{B}}=\frac{\Delta N_{\mathrm{B}}}{m_{\mathrm{an}}+m_{\mathrm{cat}}}
    \label{eq:GammaB}
\end{equation}
where $\Delta N_{\mathrm{B}}$ is the amount of stored boron in moles, $m_{\mathrm{an}}$ is the anode mass, and $m_{\mathrm{an}}$ is the cathode mass. To calculate (for the model) or measure (for the experiment) the value of $\Delta N_{\mathrm{B}}$, we follow the definition proposed by Hawks et al.\cite{Hawks2019Metrics}
\begin{equation}
    \Delta N_{i}\equiv\int_{t_{\mathrm{col}}}{A_{\mathrm{c}}v_{\mathrm{w}}\left(c_{\mathrm{F},i}-c_{\mathrm{cs},i}\right)dt}.
\end{equation}
where $t_{\mathrm{col}}$ is the collection time of effluent.
The following tables present the values used in this work. \cref{tab:general} presents the parameters used for all the simulations. \cref{tab:electrodes} presents values of the parameters which differ between the simulations. Including the simulations which were compared to experiments (\cref{fig:ElectrodesOrder,fig:Vcell}) and the simulations for theoretical analysis alone (\cref{fig:NoB_PeCfeed}). The parameters used to describe the electrodes are based on previous works.\cite{Bouhadana2011,Kim2017,Guyes2019,Uwayid2020,Guyes2021}
\begin{table}[!htbp]
    \centering
    \caption{Parameters used in all the simulations performed for this research.}
    \begin{tabular}{|l|p{3.0cm}|p{4.0cm}|}
        \hline
         \textbf{Property} & \textbf{Value} & \textbf{Description} \\
         \hline
         $l_{\mathrm{an}}=l_{\mathrm{cat}}=l_{\mathrm{e}}$ & 0.6 mm & Electrode thickness \\
         \hline
         $l_{\mathrm{sep}}$ & $\SI{65}{\micro\meter}$ & Separator thickness \\
         \hline
         $p_{\mathrm{sep}}$ & 0.8 & Separator porosity\\
         \hline
         $\tau_{\mathrm{sep}}$ & $p_{\mathrm{sep}}^{-0.5}$=1.12 & Separator tortuosity\\
         \hline
         $l_{\mathrm{res}}$ & 40$l_{\mathrm{e}}$=24 mm & Reservoir thickness\\
         \hline
         $p_{\mathrm{res}}$ & 1 & Reservoir porosity\\
         \hline
         $\tau_{\mathrm{res}}$ & $p_{\mathrm{res}}^{-0.5}$=1 & Reservoir tortuosity\\
         \hline
         $D_{\mathrm{s}}$ & $\tfrac{1}{2}\left(D_{\mathrm{Na}}^{-1}+D_{,\mathrm{Cl}}^{-1}\right)^{-1}$= =1.576 $\rm{m^2/s}$ & Salt equivalent diffusion coefficient\\
         \hline
         $t_{\mathrm{col}}$ & 10 min & Effluent collection time\\
         \hline
         $V_{\mathrm{mix}}$ & 1.86 mL & Downstream mixing volume\\
         \hline
         $A_{\mathrm{c}}$ & 6.25 $\rm{cm^2}$ & Cell cross-section area \\
         \hline
    \end{tabular}
    \label{tab:general}
\end{table}

\begin{table}[!htbp]
    \centering
    \caption{System parameters for simulations compared to experiments (\cref{fig:ElectrodesOrder,fig:Vcell}) or used only for theoretical analysis (\cref{fig:NoB_PeCfeed}).}
    \begin{tabular}{|l|p{2.5cm}|p{2.5cm}|p{3.5cm}|}
        \hline
         \textbf{Property} & \textbf{Value (\cref{fig:ElectrodesOrder,fig:Vcell})} & \textbf{Value (\cref{fig:NoB_PeCfeed})} & \textbf{Description} \\
         \hline
         $\rm{pH_F}$ & 6.3 & 7.0 & Feed pH \\
         \hline
         $c_{\mathrm{F,B}}$ & 0.37 mM & 0, 0.37 mM & Feed boron concentration \\
         \hline
         EER & 2.57 $\Omega$ & 0 $\Omega$ & Electronic resistance \\
         \hline
         $p_{\mathrm{mA}}$ & 0.698 & 0.70 & Macropores porosity \\
         \hline
         $\tau_{\mathrm{mA}}$ & $p_{\mathrm{mA}}^{-0.5}$=1.197 & $p_{\mathrm{mA}}^{-0.5}$=1.195 & Macropores tortuosity\\
         \hline
         $p_{\mathrm{mi}}$ & 0.154 & 0.172 & Micropores porosity\\
         \hline
          $C_{\mathrm{S,X}}=C_{\mathrm{S,y}}$ & 200 F/mL & 150 F/mL & Stern capacitance\\
         \hline
         $\alpha_{\mathrm{mi,X}}$ & 0.5 & 0.5 & Acidic groups volume fraction\\
         \hline
         $c_{\mathrm{chem,X,t}}$ & 0.80 M & 0.25 M & Concentration of acidic groups\\
         \hline
         $pK_{\ce{XH}}$ & 4.9 & 5.0 & Reaction's \textit{pK} in the acidic region\\
         \hline
         $\alpha_{\mathrm{mi,Y}}$ & 1-$\alpha_{\mathrm{mi,X}}$=0.5 & 1-$\alpha_{\mathrm{mi,X}}$=0.5 & Basic groups volume fraction\\
         \hline
         $c_{\mathrm{chem,Y,t}}$ & 0.60 M & 0.25 M & Concentration of basic groups\\
         \hline
         $pK_{\ce{YH+}}$ & 8.5 & 9.0 & Reaction's \textit{pK} in the basic region\\
         \hline
    \end{tabular}
    \label{tab:electrodes}
\end{table}

\subsection{Numerical model}
The equations presented in the theory subsection were solved using COMSOL multiphysics, which utilizes the finite elements method for the solution of algebraic and partial differential equations. The initial conditions were solved using a stationary study-step, coupling the equations by using a segregated solver, and multifrontal massively parallel sparse direct solver (MUMPS) as the direct solver. The transient equations were solved using a transient study-step, utilizing a fully-coupled solver where the damping factor for the nonlinear method varies between $10^{-18}$ to $10^{-4}$, and the MUMPS solver was used as the direct linear method. Moreover, the maximum element size within the cell is $l_{\mathrm{e}}/300$ and in the reservoirs is $2l_{\mathrm{e}}/15$. Last, the initial time step is $5\cdot10^{-11}\tau_{\mathrm{D}}$, for $t<0.02\tau_{\mathrm{D}}$ the maximum step-size is $4.4\cdot 10^{-5}\tau_{\mathrm{D}}$, while for later times the maximum step-size is $0.0044\tau_{\mathrm{D}}$.

\bibliographystyle{unsrt}
\bibliography{sample}

\end{document}